# Channel and Multiuser Diversities in Wireless Systems: Delay-Energy Tradeoff

Prasanna Chaporkar, Kimmo Kansanen, Ralf R. Müller

Norwegian University of Science and Technology

# Channel and Multiuser Diversities in Wireless Systems: Delay-Energy Tradeoff

**Abstract**

We consider a communication system with multi-access fading channel. Each user in the system requires certain rate guarantee. Our main contribution is to devise a scheduling scheme called "Opportunistic Super-position Coding" that satisfies the users' rate requirements. Using mean-field analysis, i.e., when the number of users go to infinity, we analytically show that the energy required to guarantee the required user rate can be made as small as required at the cost of a higher delay ("delay-energy tradeoff"). We explicitly compute the delay under the proposed scheduling policy and discuss how delay differentiation can be achieved. We extend the results to multi-band multi-access channel. Finally, all the results can be generalized in a straightforward fashion to broadcast channel due to the AWGN multiaccess-broadcast duality.

# I. INTRODUCTION

A comprehensive treatment of multi-access fading channels can be found in [1], [2]. In these papers, Tse and Hanly have characterized the so called *throughput capacity* and *delay-limited capacity* of the multi-access block fading channel with Gaussian noise assuming that perfect Channel State Information (CSI) is causally available at the transmitters and the receiver. The throughput capacity region quantifies the achievable rate region with average power constraint for ergodic fading. For the delay limited capacity, each user must be given the required rate irrespective of its fading state. The aim is to obtain a coding and power allocation scheme to minimize the energy while guaranteeing the rate in every slot. Here, *slot* refers to the time duration required to transmit a block of symbols over which the fading state remains unaltered. Thus, slot duration is the channel coherence time.

The notion of throughput capacity leads to schemes that take advantage of the differential channel qualities (*"channel diversity"*) of the users. Specifically, it has been shown that the sum rate in the system is maximized by letting only one user with the best channel to transmit. Such schemes that take current channel states into account while making scheduling decisions are referred to as *"Opportunistic Scheduling"* and may result in unfair rate allocation if the fading statistics are not symmetric. In wireless systems, the fading statistics are typically not symmetric because of many reasons that include the so called *"near-far"* effect. To alleviate this limitation, several opportunistic scheduling schemes with fairness constraints have been designed [3], [4]. Among them, Proportional Fair Scheduling (PFS) has many desirable properties including provable fairness guarantees and suitability for on-line implementation, i.e., without prior knowledge of channel statistics [5]. In spite of these desirable features, PFS suffers from two limitations. First, PFS does not provide the required rate to the users, but the rate allocation is done to maximize certain utility function and the allocated rates depend on the channel statistics of the users. Second, it does not guarantee delay as the users are scheduled in the random slots, depending on their fading states and the resource allocation in the previous slots.

As discussed above, the notion of throughput capacity is relevant for delay-tolerant data applications. On the contrary, the notion of delay-limited capacity is relevant for the applications that have the strictest delay requirement, namely, the required rate should be given to each user in each slot irrespective of the channel conditions. The delay for such schemes is always one slot. Note that these schemes cannot make use of channel diversity over time. Instead these schemes use *"multiuser diversity"*, i.e., they exploit variation in channel quality of various users in the same slot. It has been shown that super-position coding and successive decoding (aka successive cancellation) minimizes energy for achieving the required rates [2]. In this signaling scheme, all the users transmit simultaneously, and the receiver decodes in the decreasing order of the received power treating the undecoded signal as noise. One of the attractive feature of this coding strategy is that given the ordered channel state sequence, the power allocation for the optimal signaling can be obtained using a greedy procedure. This has significant impact on practical implementation.

Note that opportunistic scheduling exploits channel diversity to guarantee rates that maximize certain utility or/and guarantee fairness, while super-position coding and successive decoding exploits multiuser diversity to guarantee the required rate and the strict delay of one slot in energy efficient fashion. Many applications cannot afford to sustain indefinite delay variability in attaining the fair rate (as in PFS) without suffering significant performance penalty [6], but they also do not need the strict delay of one slot (as in delay-limited schemes), i.e., they have *limited* delay tolerance. With limited delay tolerance in multi-access channel, it is possible to exploit both channel and multi-user diversities. *Our aim is to understand how delay tolerance of the application can be exploited so as to improve the energy efficiency (delay-energy tradeoff).*

In literature, the delay-energy tradeoff is typically studied assuming no multipath fading (AWGN channel) [7], [8] or assuming non-causal CSI at transmitters and the receiver [9], [10], [11], [12]. In an AWGN channel, the problem of minimizing energy with strict deadline requirements is addressed using filter theory [7] and network calculus [8]. In [9], the authors have studied the delay-energy tradeoff for a single user in fading channel with the strict deadline (say $\tau$ slots), while in [10], [11], [12] the multi-access fading channel is considered. In all these works, optimal off-line algorithms that iteratively solve the underlying optimization problem have been developed, and heuristic algorithms for non-causal CSI have been proposed. Because of the equivalence between the multiplexing in time and frequency, the algorithms with non-causal CSI are more relevant when $\tau$ orthogonal frequency bands are available and the delay requirement is 1 slot, i.e., the delay limited capacity of multi-band multi-access system. In their seminal work, Cheng and Verdú have obtained the capacity region for the vector Gaussian multi-access channel [13]. Here, authors have considered the scalar Gaussian channel with ISI, which reduces to the case of independent parallel memoryless Gaussian channels through Karhunen-Loéve expansion. No multi-path fading was assumed. Yu et al. and Médard have obtained the sum capacity region of the multi-band multi-access fading channel with white Gaussian noise [14], [15]. Perfect CSI at the transmitter and the receiver is assumed in all these works. The capacity achieving power allocation is typically multi-user waterfilling scheme. The exact characterization of this multi-user waterfilling is considered difficult, and only iterative algorithms and their convergence properties are known [14], [15]. The problem of providing the desired throughput to each of the users has been addressed in multi-band multi-access fading channel with white Gaussian noise [16], [17], [18]. The aim is to obtain approximate multi-user waterfilling solution efficiently.

In practice, CSI is only available causally. In spite of this obvious limitation, the case with non-causal CSI has been studied in literature because of the following two reasons. First, the optimum schemes with non-causal CSI (*off-line* schemes) provide a benchmark for all the other schemes that guarantee the required delay. Also, the structural properties of the optimal off-line schemes facilitate valuable insights for designing near-optimal *on-line* schemes (schemes with causal CSI). Second, in the fixed delay case, analytically it is difficult, if not impossible, to design optimal on-line schemes barring the trivial cases in which future fading states can be accurately predicted. This can be seen as follows. Since the required delay $\tau$ is finite, any fading realization is possible with positive probability. Thus, for any given on-line scheme, one can play the role of an adversary and orchestrate future fading states so as to make the scheme suboptimal compared to optimal off-line scheme. In view of these reasons, instead of the fixed delay constraint, average delay constraint is considered while designing optimal on-line schemes [19], [20]. From the quality of service point of view, guaranteeing the fixed delay is more desirable than guaranteeing the average delay for the real-time applications. But, if the delay variance is small, then the higher layers can employ mechanisms like playback buffer to cope up with the delay variability. In [19], [20], a single user fading channel is considered, and the minimization of average delay for the given average power constraint has been addressed using the framework of Markov Decision Processes (MDP). Again, only the structural properties of the optimal on-line scheme have been derived except in certain special cases. The exact on-line scheme has to be obtained by solving Bellman's equations, which is computationally expensive. The results in [19], [20] for a single user case are extended to multi-access channel by Neely [21]. Here, the author has considered the mean system delay, i.e., the average delay over all the users. We note that guaranteeing the mean system delay does not guarantee the required delay performance to individual users. In fact, such schemes tend to favor users with a better channel by providing smaller delays to these at the expense of the users with a worse channel and still maintaining the desired mean system delay. Our aim is to quantify delay-energy tradeoff while guaranteeing average delay to each of the users. In queueing theory literature, frameworks for designing scheduling schemes that minimize certain utility (energy in our case) while guaranteeing bounded mean delay [22], [23], [24], [25]. These frameworks guarantee the bounded delay, but do not guarantee the desired delay to each of the users as we do.

From the above discussion, it should be clear that the closed form expressions for the minimum energy required for guaranteeing the desired delay are not available for both on-line and off-line schemes. Thus, the energy-delay tradeoff has to quantified using numerical evaluations for specific scenarios of interest. Our aim is to obtain closed form expressions to the quantify delay-energy tradeoff for average delay constraint in multi-band multi-access fading channels with causal CSI at the transmitters and the receiver. We consider the multi-band systems as they can model many practical systems including OFDM systems, channels with frequency selective fading and ISI channels. In our analysis, we allow for general fading distribution, and also consider random arrivals of the symbols from the higher layer into physical layer buffer to model real-time applications in which the symbols are generated in real-time. For analytical tractability, we use the following approach: we design a parametrized scheduling policy called *"Opportunistic Super-position Coding" (OSPC)* that exploits the channel diversity by scheduling a set users with high channel gains only, and among these users it uses super-position coding and successive decoding to exploit multiuser diversity. One of the main challenges in designing such schemes is the quantification of its performance. The quantification allows for the optimal and guaranteed control. We explicitly quantify the per user delay and the total energy requirement for the proposed policy. Thus, given the delay requirement, we can efficiently choose the appropriate parameter values so as to minimize the energy while guaranteeing the required delay. Using numerical computations, we show that a small delay tolerance can be exploited to achieve the significant energy savings. We also compare the performance of the proposed policy with PFS and the delay-limited schemes.

The paper is arranged as follows. In Section II, we present our system model. In Section III, we describe the OSPC policy, and in Section IV obtain analytical guarantees. In Section V, we discuss extensions of OSPC to multi-band multi-access channel and to provide delay differentiation. In Section VI, we compare the performance of OSPC with PFS and delay-limited schemes using numerical computations. Finally, in Section VII, we conclude.

## II. System Model

We consider a multi-access system with $K$ users that are placed at random in a cell. Time is slotted. Each user $i$ requires a certain fraction of the total rate provided in a system, i.e., the required rate $R = \frac{\Gamma}{K}$, where $\Gamma$ denotes the spectral efficiency. Alternatively, $R_i(t) = \frac{\Gamma}{K}\nu_i(t)$ denotes the arrivals for user $i$ in slot $t$, where $\nu_i(t)$ is a random variable (r.v.). We assume that all the moments for $\nu_i(t)$ are finite and $\mathbb{E}[\nu_i(t)] = 1$ for every $i$. Note that random variables with finite support have all the moments finite. In networks, the arrival rate is typically limited by the link capacities which may be large but finite. Moreover, we assume that the arrivals are independent and identically distributed (i.i.d.) across both slots and users. The arrivals are queued into an infinite buffer before served. Without loss of generality, let the system start in slot 1. Hence, users' buffers are empty at the beginning of slot 1. We primarily discuss the uplink communication (multi-access channel), but the results can be generalized in a straightforward fashion for the downlink case (broadcast channel) using Gaussian multiaccess-broadcast duality [26].

Now, we describe the model for the multi-access channel. Each user $i$ experiences the channel $d_i(t)$ in slot $t$. The channel $d_i(t)$ arises due to two independent effects, namely, *path loss* (denoted by $s_i$) and *short term fading* (denoted by $f_i(t)$). The path loss is a function of the distance between the transmitter-receiver pair. Typically, the distance between transmitter and receiver changes very slowly with respect to the signal bandwidth. Hence, we assume that the path loss is constant from slot to slot given the distance of a user from the transmitter. On the contrary, $f_i(t)$ depends on the scattering environment around the user and changes in time depending on the channel Doppler bandwidth. We assume that $f_i(t)$ changes from slot to slot and is i.i.d. across both users and slots. Now, $d_i(t) = s_i f_i(t)$. This is referred to as the block fading model [27]. Let $E_i^R(t)$ ($E_i(t)$, resp.) denote the received (transmitted, resp.) energy from user $i$ in slot $t$. Then, $E_i^R(t) = d_i(t)E_i(t)$. Let $\vec{d}(t) = [d_1(t) \ \ldots \ d_K(t)]$. Note that the fading for users is *not* symmetric. The distribution of pathloss for a randomly placed user is denoted by $\Psi(\cdot)$, i.e., $\mathbb{P}\{S \leq x\} = \Psi(x)$ where $S$ is a generic r.v. indicating the pathloss of a randomly placed user. Note that $\Psi(\cdot)$ depends on the distribution of the users' placement and the path loss function. Also, let $\Phi(\cdot)$ denote the short term fading distribution, i.e., $\mathbb{P}\{f_i(t) \leq x\} = \Phi(x)$ for every user $i$ and slot $t$. Let $N_0$ denote the noise power spectral density.

*Definition 1 (Scheduling Policy):* A *scheduling policy* $\Delta$ is an algorithm that in each slot $t$ determines the rate vector $\vec{\rho}^\Delta(t) = [\rho_1^\Delta(t) \ \ldots \ \rho_K^\Delta(t)]$, and serves each user $i$ with rate $\rho_i^\Delta(t)$.

We assume that the perfect CSI is available at the receivers and the transmitter in every slot $t$. Thus, a scheduling policy may adapt $\vec{\rho}^\Delta(t)$ to the channel state.

Now, we discuss the power allocation for each user in order to achieve the desired rates $\vec{\rho}$. Fix a power allocation given by a vector $\vec{E} = [E_1 \ \cdots \ E_K]$, where $E_i$ denote the power of $i^{\text{th}}$ user. Ahleswede [28] and Liao [29] have shown that the capacity region of the Gaussian multi-access channel with the time invariant fading $\vec{d}$ for the power allocation $\vec{E}$ is the set of rate vectors $\vec{\rho}$ that satisfy

$$\sum_{i \in \mathcal{S}} \rho_i \leq \log\left(1 + \frac{\sum_{i \in \mathcal{S}} d_i E_i}{N_0}\right),$$

for every $\mathcal{S} \subseteq \{1, \ldots, K\}$. The rates are achieved with the standard random Gaussian code book with variance $E_i$ for user $i$. The signaling is as follows: all the users transmit simultaneously (superposition coding) and the receiver decodes successively in the decreasing order of the received power. From this result, it is straight forward to see that a given rate vector $\vec{\rho}$ is feasible with power allocation $\vec{E}$ in Gaussian multi-access channel with fading $\vec{d}$ if and only if (iff) there exists a permutation $\pi_1, \ldots, \pi_K$ of $\{1, \ldots, K\}$ such that for every $i$

$$E_{\pi_i} \geq \frac{N_0}{d_{\pi_i}} \exp\left(\sum_{k < i} \rho_{\pi_k}\right) [\exp(\rho_{\pi_i}) - 1].$$

Note that there may exist many power allocations that can realize the rate vector $\vec{\rho}$. Since we aim to obtain power efficient strategies, we seek an optimal power allocation that minimizes the sum power while achieving the desired rates. Such power allocation can be explicitly obtained as follows [1]. Let $\vec{\pi}$ denote the permutation such that $d_{\pi_1} \leq d_{\pi_2} \leq \cdots \leq d_{\pi_K}$. Then, the power allocation given below is optimal.

$$E_{\pi_i} = \frac{N_0}{d_{\pi_i}} \exp\left(\sum_{k < i} \rho_{\pi_k}\right) [\exp(\rho_{\pi_i}) - 1], \tag{1}$$

for every $i \in \{1, \ldots, K\}$. Note that for the optimal signaling, the successive decoding order depends only on channel gains, but not on rates. We assume that once a scheduling policy $\Delta$ determines the rate allocation, the corresponding power allocation is done as per (1).

We consider three performance measures, namely, stability, energy efficiency and delay. Next, we formally define these.

*Definition 2 (Busy Period):* A *busy period* for user $i$ is the set of consecutive slots in which its queue length is greater than zero.

*Definition 3 (Stability):* Let $\mathcal{B}_{u,i}$ denote the length of the $u^{\text{th}}$ busy period of user $i$. The system is said to be *strongly stable* if for every user $i$

$$\limsup_{U \to \infty} \frac{1}{U} \sum_{u=1}^{U} \mathcal{B}_{u,i} < \infty.$$

*Definition 4 (Energy Efficiency):* We define the energy efficiency in slot $t$ as

$$\left( \frac{E_b(t)}{N_0} \right)_{\text{sys}} \overset{\text{def}}{=} \frac{\sum_{i=1}^{K} E_i(t)}{N_0 \Gamma}.$$

Then, the energy efficiency is defined as

$$\left( \frac{E_b}{N_0} \right)_{\text{sys}} \overset{\text{def}}{=} \limsup_{T \to \infty} \frac{1}{T} \sum_{t=1}^{T} \left( \frac{E_b(t)}{N_0} \right)_{\text{sys}}.$$

*Definition 5 (Delay):* Delay for the $u^{\text{th}}$ arrival of user $i$ (denoted by $D_{u,i}$) is the number slots between its departure and arrival. The delay for user $i$ is then defined as

$$D_i \overset{\text{def}}{=} \limsup_{U \to \infty} \frac{1}{U} \sum_{u=1}^{U} D_{u,i}.$$

In general, we augment the notation to denote the dependence of various quantities on the scheduling policy $\Delta$ by using $\Delta$ as a superscript, e.g., $D_i^{\Delta}$ will indicate the delay for user $i$ under policy $\Delta$.

## III. Opportunistic Super-position Coding ($\Delta^*$)

The scheduling policy $\Delta^*$ is parametrized by a variable $\kappa$. To specify this dependence, we denote the OSPC scheduling policy as $\Delta^*(\kappa)$. The scheduling decisions are taken as follows in every slot $t$.

- Select all users $i$ such that $f_i(t) > \kappa$.
- Allocate rate $\rho_i(t)$ to each of the chosen users $i$ so that everything in its buffer is served, $\rho_i(t) = 0$ for others.

Note that $\Delta^*(\kappa)$ selects users based on $f_i(t)$ and not on $d_i(t)$. This is for the reasons of fairness as $f_i(t)$ has the same distribution for every user, while the distribution of $d_i(t)$ depends on the distance of user $i$ from the base station and is biased towards stronger channels the closer the user is to the base station. Moreover, since the pathloss is constant for each user, it suffices to determine rate allocation in each slot depending only on short term fading which is time varying, and thus, unlike pathloss, provides diversity which can be exploited.

We refer to $\kappa$ as *opportunism threshold* as it dictates how opportunistic OSPC is in exploiting the channel diversity. The opportunism threshold plays a key role in determining delay and energy consumption in the system. We intuitively explain how. Note that appropriate choice of $\kappa$ allows for eliminating users that are in deep fade in a given slot. Typically, few users with very bad channel dominate the total energy consumption in the system for the delay-limited schemes. Thus, even a small value of the opportunism threshold should significantly improve energy efficiency of the system. Specifically, we can expect the energy consumption to decrease monotonically with increase in $\kappa$. But, note that as $\kappa$ increases, each user is scheduled less frequently. Thus, for satisfying the rate requirements, the rate given to the scheduled users increases monotonically with $\kappa$. And providing the higher rate requires higher energy. In fact, for the fixed noise power, the required energy increases exponentially with increase in the rate. Summarizing, the opportunism threshold allows us to save energy by eliminating the users in deep fade, but requires higher energy to serve scheduled users. Thus, it is not clear if the opportunism threshold should improve the system performance. It, however, turns out that the energy consumption decreases monotonically as a function of $\kappa$. In other words, the energy savings caused by eliminating the worst users is more than the increase in the energy consumption to provide the higher rates to the scheduled users.

As discussed above, the users are scheduled less frequently when $\kappa$ is large. So, clearly, the delay increases monotonically with increase in $\kappa$. Thus, the opportunism threshold provides a way to achieve delay-energy tradeoff. Now, the main challenge is to quantify the delay and energy as a function of $\kappa$, so as to choose the optimal $\kappa$ that guarantees the required delay while minimizing the energy.

## IV. ANALYTICAL GUARANTEES

In this section, we obtain analytical guarantees for $\Delta^*(\kappa)$. In Theorem 1, we show that $\Delta^*(\kappa)$ strongly stable. In Theorem 2, we quantify per user delay under $\Delta^*(\kappa)$. In Theorem 3, we quantify the energy efficiency of $\Delta^*(\kappa)$. Finally, in Theorems 4 and 5 we show that the energy efficiency of $\Delta^*(\kappa)$ decreases monotonically with $\kappa$. Finally, in Theorem 5, we show that as delay goes to $\infty$, the energy required to stabilize the system becomes equal to the minimum energy required to provide the desired rates to each of the users in long term. In Subsection IV-A, we discuss the implications of these results. First, let $\gamma \stackrel{\text{def}}{=} \mathbb{P}\{f_i > \kappa\}$.

*Theorem 1 (Strong Stability):* For every $\kappa$ such that $\gamma > 0$, $\Delta^*(\kappa)$ is strongly stable w.p. 1.

*Proof:* Fix any user $i$. Since $\{f_i(t)\}_{t \geq 1}$ is a sequence of i.i.d. random variables, the user is scheduled is each slot w.p. $\gamma$ under $\Delta^*(\kappa)$. Moreover, each time the user is scheduled, $\Delta^*(\kappa)$ serves everything in its buffer. Also, since the arrivals are i.i.d. in slots, clearly, the busy periods are geometrically distributed i.i.d. random variables with mean $1/\gamma$. Thus, by the Strong Law of Large Numbers (SLLN) for $\gamma > 0$

$$\limsup_{U \to \infty} \frac{1}{U} \sum_{u=1}^{U} \mathcal{B}_{u,i} = \lim_{U \to \infty} \frac{1}{U} \sum_{u=1}^{U} \mathcal{B}_{u,i} = \frac{1}{\gamma} \quad \text{w.p. 1.}$$

Now, the result follows from Definition 3 as $\gamma > 0$. ∎

*Theorem 2 (User Delay):* The delay for any user $i$ under $\Delta^*(\kappa)$ is

$$D_i^{\Delta^*(\kappa)} = \frac{1}{\gamma} \quad \text{w.p. 1.} \tag{2}$$

*Proof:* Let $\{T_n\}_{n \geq 1}$ denote the sequence of time slots such that $f_i(T_n) \geq \kappa$ and $f_i(t) < \kappa$ for every other $t$. In words, $T_n$ denotes the $n^{\text{th}}$ time slot in which user $i$ is chosen by $\Delta^*(\kappa)$. Since $\{f_i(t)\}_{t \geq 1}$ is a sequence of i.i.d. random variables, $\{(T_{n+1} - T_n)\}_{n \geq 1}$ is also a sequence of i.i.d. random variables with geometric distribution and mean $1/\gamma$. Here, $T_0 = 0$. Let $U_n$ denote the total number of arrivals of user $i$ until $T_n$. Note that $(T_{n+1} - T_n)$ and arrivals in each slot are i.i.d., and fading and arrivals are mutually independent. Hence, we conclude that $\{(U_n - U_{n-1})\}_{n \geq 1}$ is a sequence of i.i.d. random variables. Here, $U_0 = 0$. By SLLN it follows that

$$\begin{aligned}
\lim_{n \to \infty} \frac{U_n}{n} &= \mathbb{E}[U_1] \quad \text{w.p. 1} \\
&= \mathbb{E}[T_1] \mathbb{E}[R_i(1)] \quad \text{w.p. 1} \\
&= \frac{\Gamma}{\gamma K} \quad \text{w.p. 1.}
\end{aligned} \tag{3}$$

Let $\overline{D}_{n,i}$ denote the sum of the delays experienced by all the arrivals in duration $(T_{n-1}, T_n]$. Again notice that $\{\overline{D}_{n,i}\}_{n \geq 1}$ is an i.i.d. sequence. Furthermore, the arrivals in slot $t \in (T_{n-1}, T_n]$ experience delay of $(T_n - t + 1)$ slots. Thus,

$$\overline{D}_{n,i} = \sum_{t=T_{n-1}+1}^{T_n} (T_n - t + 1) R_i(t).$$

Note that because of the independence between i.i.d. sequences $\{T_n\}_{n \geq 1}$ and $\{R_i(t)\}_{t \geq 1}$,

$$\mathbb{E}[\overline{D}_{n,i} \mid T_{n-1}, T_n] = \frac{\Gamma}{K} \frac{(T_n - T_{n-1})(T_n - T_{n-1} + 1)}{2}.$$

Thus, we obtain

$$\begin{aligned}
\mathbb{E}[\overline{D}_{n,i}] &= \frac{\Gamma}{2K} \mathbb{E}[T_1(T_1 + 1)] \\
&= \frac{\Gamma}{K\gamma^2}.
\end{aligned}$$

Moreover, by SLLN,

$$\lim_{n \to \infty} \frac{\overline{D}_{n,i}}{n} = \frac{\Gamma}{K\gamma^2} \quad \text{w.p. 1.} \tag{4}$$

Now, we evaluate $\limsup_{U \to \infty} \frac{1}{U} \sum_{u=1}^{U} D_{u,i}$. First consider the subsequence $\left\{ \frac{1}{U_n} \sum_{u=1}^{U_n} D_{u,i} \right\}_{n \geq 1}$.

$$\frac{1}{U_N} \sum_{u=1}^{U_N} D_{u,i} = \frac{1}{U_N} \sum_{n=1}^{N} \overline{D}_{n,i}.$$

Thus,

$$
\begin{aligned}
\lim_{N \to \infty} \frac{1}{U_N} \sum_{u=1}^{U_N} D_{u,i} &= \lim_{N \to \infty} \frac{1}{U_N} \sum_{n=1}^{N} \overline{D}_{n,i} \\
&= \lim_{N \to \infty} \frac{N}{U_N} \lim_{N \to \infty} \frac{1}{N} \sum_{n=1}^{N} \overline{D}_{n,i} \\
&= \frac{1}{\gamma} \quad \text{w.p. 1.}
\end{aligned}
$$

(5)

The last equality follows from (3) and (4).

Now, consider any $U \in \{U_{n-1}+1, \ldots, U_n\}$ and note that

$$\frac{1}{U_N} \sum_{n=1}^{N-1} \overline{D}_{n,i} \leq \frac{1}{U} \sum_{u=1}^{U} D_{u,i} \leq \frac{1}{U_{N-1}} \sum_{n=1}^{N} \overline{D}_{n,i}.$$

(6)

Note that as $U \to \infty$, $N \to \infty$, and $\lim_{N \to \infty} \frac{U_N}{U_{N-1}} = 1$. Now, by taking limit $U \to \infty$ in (6), we conclude from (5) that

$$\lim_{U \to \infty} \frac{1}{U} \sum_{u=1}^{U} D_{u,i} = \frac{1}{\gamma} \quad \text{w.p. 1.}$$

Thus, the result follows from Definition 5. ∎

*Note:* Users' delays do not depend on the distribution of their arrival processes given that the mean is finite. Moreover, the delays for the users are not correlated. Thus, for a given $\kappa$, users may have different distributions for their respective arrival processes and yet receive the same delay as long as these processes are independent. Furthermore, user's delay guarantee is independent of the number of users in the system.

Let $\overline{F}_\kappa(\cdot)$ denote the fading distribution of user $i$ who is randomly placed in a cell given that $f_i(t) > \kappa$, i.e., $\overline{F}_\kappa(x) = \mathbb{P}\{s_i f_i(t) \leq x | f_i(t) > \kappa\}$, where $s_i$ is a r.v. denoting the path loss of a randomly placed user $i$. Note that $\overline{F}_\kappa(\cdot)$ does not depend on time as the short term fading is i.i.d. across the slots. Now, let $\mathcal{A}_K^\kappa(x,t)$ denote the set of users that are chosen for service under $\Delta^*(\kappa)$ in slot $t$ such that their channel gains are less than or equal to $x$, i.e., $\mathcal{A}_K^\kappa(x,t) = \{i \in \{1, \ldots, K\} : f_i(t) > \kappa \text{ and } d_i(t) \leq x\}$. Clearly, $\mathcal{A}_K^\kappa(\infty,t)$ denotes the set of all the users chosen by $\Delta^*(\kappa)$ in slot $t$. Let $|\mathcal{A}_K^\kappa(x,t)|$ denote the cardinality of $\mathcal{A}_K^\kappa(x,t)$.

Next, we obtain the energy efficiency of $\Delta^*(\kappa)$. But, before that we state two results that we use to prove the required.

*Lemma 1:* For every $t$,

$$\lim_{K \to \infty} \sup_x \left| \sum_{k \in \mathcal{A}_K^\kappa(x,t)} \rho_k^{\Delta^*(\kappa)}(t) - \Gamma \overline{F}_\kappa(x) \right| = 0 \quad \text{w.p. 1.}$$

(7)

*Proof:* The proof is presented in Appendix A. ∎

*Lemma 2:* For every $t$ and $x$,

$$\lim_{K \to \infty} \sup_{i \in \mathcal{A}_K^\kappa(x,t)} \rho_i^{\Delta^*(\kappa)}(t) = 0 \quad \text{w.p. 1.} \tag{8}$$

*Proof:* The proof is presented in Appendix B. ∎

*Theorem 3 (Energy Efficiency):* In the mean field, i.e. as $K \to \infty$, the $(E_b/N_0)_{\text{sys}}$ under policy $\Delta^*(\kappa)$ is given by

$$
\begin{aligned}
\left(\frac{E_b}{N_0}\right)_{\text{sys}}^{\Delta^*(\kappa)} &= \mathbb{E}_{\overline{F}_\kappa}\left[\frac{e^{\Gamma \overline{F}_\kappa(x)}}{x}\right] \\
&= \int_0^\infty \frac{1}{x} \exp(\Gamma \overline{F}_\kappa(x)) \mathbf{d}\overline{F}_\kappa(x) \quad \text{w.p. 1.}
\end{aligned}
$$

*Proof:* Fix arbitrary slot $t$. We show that under $\Delta^*(\kappa)$, $\left(\frac{E_b(t)}{N_0}\right)_{\text{sys}}^{\Delta^*(\kappa)}$ is the same in every slot $t$ as $K \to \infty$. So, for convenience, we drop $t$ in the notation.

Now, from (1) and Definition 4, it follows that

$$\left(\frac{E_b}{N_0}\right)_{\text{sys}}^{\Delta^*(\kappa)} = \sum_{i \in \mathcal{A}_K^\kappa(\infty)} \frac{1}{\Gamma d_i} e^{\sum_{k \in \mathcal{A}_K^\kappa(d_i)} \rho_k^{\Delta^*(\kappa)}} (e^{\rho_i^{\Delta^*(\kappa)}} - 1). \tag{9}$$

Since $\exp(\cdot)$ is a continuous function, Lemma 1 implies that there exists a sequence of non-negative r.v.'s $\{\epsilon_K\}_{K \geq 1}$ independent of $d_i$ such that $\lim_{K \to \infty} \epsilon_K = 0$ w.p. 1, and

$$e^{-\epsilon_K} e^{\Gamma \overline{F}_\kappa(d_i)} \leq e^{\sum_{k \in \mathcal{A}(d_i)} \rho_k^{\Delta^*(\kappa)}} \leq e^{\epsilon_K} e^{\Gamma \overline{F}_\kappa(d_i)}. \tag{10}$$

Thus from (9) and (10),

$$\frac{e^{-\epsilon_K}}{\Gamma} \sum_{i \in \mathcal{A}_K^\kappa(\infty)} \frac{1}{d_i} e^{\Gamma \overline{F}_\kappa(d_i)} (e^{\rho_i^{\Delta^*(\kappa)}} - 1) \leq \left(\frac{E_b}{N_0}\right)_{\text{sys}}^{\Delta^*(\kappa)} \leq \frac{e^{\epsilon_K}}{\Gamma} \sum_{i \in \mathcal{A}_K^\kappa(\infty)} \frac{1}{d_i} e^{\Gamma \overline{F}_\kappa(d_i)} (e^{\rho_i^{\Delta^*(\kappa)}} - 1). \tag{11}$$

Now, by Lemma 2, we conclude for a large enough $K$ that

$$(e^{\rho_i^{\Delta^*(\kappa)}} - 1) \approx \rho_i^{\Delta^*(\kappa)} = \frac{\Gamma \overline{\nu}_i}{K}.$$

Note that the approximation becomes tighter as $K$ becomes larger. Thus,

$$\frac{e^{-\epsilon_K}}{\Gamma} \sum_{i \in \mathcal{A}_K^\kappa(\infty)} \frac{e^{\Gamma \overline{F}_\kappa(d_i)}}{d_i} \left(\frac{\Gamma \overline{\nu}_i}{K}\right) \leq \left(\frac{E_b}{N_0}\right)_{\text{sys}}^{\Delta^*(\kappa)} \leq \frac{e^{\epsilon_K}}{\Gamma} \sum_{u=1}^\infty \sum_{i \in \mathcal{A}_K^\kappa(\infty)} \frac{e^{\Gamma \overline{F}_\kappa(d_i)}}{d_i} \left(\frac{\Gamma \overline{\nu}_i}{K}\right). \tag{12}$$

Let us consider the following term in (12),

$$
\begin{aligned}
&\lim_{K \to \infty} e^{\epsilon_K} \sum_{i \in \mathcal{A}_K^\kappa} \frac{e^{\Gamma \overline{F}_\kappa(d_i)}}{d_i} \frac{\overline{\nu}_i}{K} \\
&= \lim_{K \to \infty} e^{\epsilon_K} \lim_{K \to \infty} \frac{|\mathcal{A}_K^\kappa(\infty)|}{K} \lim_{|\mathcal{A}_K^\kappa(\infty)| \to \infty} \frac{1}{|\mathcal{A}_K^\kappa(\infty)|} \sum_{i \in \mathcal{A}_K^\kappa(\infty)} \frac{e^{\Gamma \overline{F}_\kappa(d_i)}}{d_i} \overline{\nu}_i \\
&= \gamma \mathbb{E}\left[\frac{e^{\Gamma \overline{F}_\kappa(d_i)}}{d_i}\right] \mathbb{E}[\overline{\nu}_i] \quad \text{w.p. 1.}
\end{aligned}
\tag{13}
$$

Similarly,

$$\lim_{K \to \infty} e^{-\epsilon_\kappa} \sum_{i \in \mathcal{A}_K^\kappa(\infty)} \frac{e^{\Gamma \overline{F}_\kappa(d_i)}}{d_i} \frac{\overline{\nu}_i}{K} = \gamma \mathbb{E}\left[\frac{e^{\Gamma \overline{F}_\kappa(d_i)}}{d_i}\right] \mathbb{E}[\overline{\nu}_i] \text{ w.p. } 1. \qquad (14)$$

The first expectation in (13) and (14) is with respect to the distribution $\overline{F}_\kappa(\cdot)$, while the second is with respect to the distribution of the arrival process. The relations (13) and (14) hold because the channel gains $d_i$'s of the chosen users can be viewed as i.i.d. variables drawn from the distribution $\overline{F}_\kappa(\cdot)$. This can be seen as follows. Since $f_k$'s are i.i.d. irrespective of the distance between the user and receiver, the scheduling decision can be viewed as scheduling each user w.p. $\gamma$ independently. Since the users are placed at random, $s_k$ is a deterministic function of distance, and $f_k$ and $s_k$ are independent, we conclude that $d_k$'s for the chosen users are i.i.d. and each $d_k$ is distributed as $\overline{F}_\kappa(\cdot)$. Thus, $\left\{\frac{e^{\Gamma \overline{F}_\kappa(d_i)}}{d_i}\right\}_{i=1,\ldots,|\mathcal{A}_K^\kappa(\infty)|}$ is an i.i.d. sequence. Thus, from (13) and (14)

$$\begin{aligned}
\lim_{K \to \infty} e^{\epsilon_\kappa} \sum_{i \in \mathcal{A}_K^\kappa(\infty)} \frac{e^{\Gamma \overline{F}_\kappa(d_i)}}{d_i} \frac{\overline{\nu}_i}{K} &= \lim_{K \to \infty} e^{-\epsilon_\kappa} \sum_{i \in \mathcal{A}_K^\kappa(\infty)} \frac{e^{\Gamma \overline{F}_\kappa(d_i)}}{d_i} \frac{\overline{\nu}_i}{K} \\
&= \int_0^\infty \frac{1}{x} \exp(\Gamma \overline{F}_\kappa(x)) \mathbf{d}\overline{F}_\kappa(x) \text{ w.p. } 1. \qquad (15)
\end{aligned}$$

Thus the required follows from Definition 4. ∎

*Note:* The energy efficiency does not depend on the distribution of the users' arrival processes, but depends only on the mean as long as the processes are independent and have all the moments finite.

*Theorem 4 (Monotonicity of Energy Efficiency):* In the mean field for every $\kappa' < \kappa$,

$$\left(\frac{E_b}{N_0}\right)_{\text{sys}}^{\Delta^*(\kappa)} \leq \left(\frac{E_b}{N_0}\right)_{\text{sys}}^{\Delta^*(\kappa')} \text{ w.p. } 1.$$

Moreover, the decrease in energy efficiency is $O\left(\frac{1}{\kappa}\right)$. Specifically,

$$\left(\frac{E_b}{N_0}\right)_{\text{sys}}^{\Delta^*(\kappa)} \leq \frac{1}{\kappa} \mathbb{E}_\Psi\left[\frac{e^{\Gamma \Psi(S)}}{S}\right] \leq \frac{e^\Gamma \mathbb{E}_\Psi[1/S]}{\kappa} \text{ w.p. } 1. \qquad (16)$$

First, we provide the intuition behind the result and then provide the proof.

*Intuition:* Note the following property of the $\log(\cdot)$ function. Let $\{a_i\}_{i \geq 1}$ denote any sequence of the non-negative real numbers and let $Z$ be any constant. Then,

$$\sum_{i=1}^\infty \log\left(1 + \frac{a_i}{Z + \sum_{u < i} a_u}\right) = \log\left(1 + \frac{\sum_{i=1}^\infty a_i}{Z}\right). \qquad (17)$$

Thus, under super position coding and successive decoding, we obtain

$$\sum_{i \in \mathcal{A}_\kappa(\infty)} \rho_i(t) = \log\left(1 + \frac{\sum_{i \in \mathcal{A}_\kappa(\infty)} E_i^R(t)}{N_0}\right), \qquad (18)$$

where $\rho_i(t)$ denote the rate requirement of user $i$ in slot $t$ and $\mathcal{A}_\kappa(\infty)$ denote the number of scheduled users scheduled under $\Delta^*(\kappa)$.

Thus, from (18) we conclude that if the sum rate to be provided is the same, then the required sum received energy at the receiver remains the same and is independent of the individual rates. Now, note that the sum rate to be provided in every slot under $\Delta^*(\kappa)$ is equal to $\Gamma$ in the mean field for every $\kappa$. This can be seen as follows.

$$
\begin{aligned}
\lim_{K\to\infty}\sum_{i\in\mathcal{A}_\kappa(\infty)}\rho_i(t) &= \lim_{K\to\infty}\sum_{i\in\mathcal{A}_\kappa(\infty)}\frac{\Gamma}{K}\overline{\nu}_i(t)\\
&= \Gamma\lim_{K\to\infty}\frac{|\mathcal{A}_\kappa(\infty)|}{K}\lim_{|\mathcal{A}_\kappa(\infty)|\to\infty}\frac{1}{|\mathcal{A}_\kappa(\infty)|}\sum_{i\in\mathcal{A}_\kappa(\infty)}\overline{\nu}_i(t)\\
&= \Gamma \quad \text{w.p. 1.}
\end{aligned}
\tag{19}
$$

Thus, it follows that the required sum received energy under $\Delta^*(\kappa)$ is the same in every slot independent of $\kappa$. Now, the required sum transmit energy depends on the channel states. Since as $\kappa$ increases, only the users with a larger channel gains are selected, we intuitively expect the sum transmit energy required to achieve the given sum received energy to decrease monotonically. Now, we prove Theorem 4. First, we state a lemma that we use to prove the theorem.

*Lemma 3:* Let $\mathcal{A}$ denote an ordered countable set of users, where the ordering is as per fading, i.e., $d_i \leq d_j$ whenever $i < j$. Let $\vec{\rho}$ and $\vec{\rho}'$ denote two different rate requirement vectors satisfying

$$
\sum_{i=1}^{k}\rho_i \geq \sum_{i=1}^{k}\rho_i' \quad \text{for every } k,
$$

Moreover, let $\vec{E}$ and $\vec{E}'$ denote the power allocation as per (1) to realize $\vec{\rho}$ and $\vec{\rho}'$, respectively. Then, $\sum_{i\in\mathcal{A}}E_i \geq \sum_{i\in\mathcal{A}}E_i'$.

*Proof:* The proof is presented in Appendix C. ∎

Now, we prove Theorem 4.

*Proof:* Let us consider two identical copies of a sample path, i.e., arrivals and fading for each of the users are the same in every slot. On the first sample path, the users are served as per $\Delta^*(\kappa)$, and on the second they are served as per $\Delta^*(\kappa')$, where $\kappa < \kappa'$. Note that $\mathcal{A}_K^{\kappa'}(x,t) \subseteq \mathcal{A}_K^{\kappa}(x,t)$ for every $K$, $x$ and $t$ as $f_i(t) > \kappa'$ implies $f_i(t) > \kappa$.

Now, fix $t$ and $K$ and define $\mathcal{A} = \mathcal{A}_K^{\kappa}(\infty,t)$. Moreover, let $\vec{\rho}(t) = \{\rho_i^{\Delta^*(\kappa)}(t) : i \in \mathcal{A}\}$ and $\vec{\rho}'(t) = \{\rho_i^{\Delta^*(\kappa')}(t)\mathbf{1}_{\{i\in\mathcal{A}_K^{\kappa'}(\infty,t)\}} : i \in \mathcal{A}\}$. Now, we show that for every $i \in \mathcal{A}$, $\sum_{k<i}\rho_k \geq \sum_{k<i}\rho_k'$. By Lemma 1, $\sum_{k<i}\rho_k \to \Gamma\overline{F}_\kappa(d_i)$ and $\sum_{k<i}\rho_k' \to \Gamma\overline{F}_{\kappa'}(d_i)$ w.p. 1 as $K \to \infty$. Moreover, for a randomly placed user $i$, $\{d_i \leq x|f_i > \kappa\} \supseteq \{d_i \leq x|f_i > \kappa'\}$ for every $x$ whenever $\kappa \leq \kappa'$. Thus, by the monotonicity of the probability measure, $\overline{F}_{\kappa'}(x) \leq \overline{F}_\kappa(x)$ for every $x$. Thus, as $K \to \infty$, $\sum_{k<i}\rho_k \geq \sum_{k<i}\rho_k'$ on any non-trivial sample path. Thus, monotonicity property follows from Lemma 3 as $t$ is arbitrary.

Now, we show the first inequality in (16). The inequality follows by observing that the fading of every chosen user $i$ satisfies $d_i \geq \kappa s_i$. We consider power allocation with these worse fading states $(\kappa s_i)$ for every chosen user. Now, observe that with this modified fading process, the randomness remains only in the pathloss $s_i$. Hence, the required follows using exactly the same arguments as that in Theorem 3. We, however, would like to mention that the inequality does not follow from the expressions stated in the statement of Theorem 3 as there the users are ordered as per $\vec{d}$ which may be different than that as per the pathloss considered here. Now, the second inequality in (16) follows by observing that $\Psi(x) \leq 1$ for every $x$ in the first inequality in (16). ∎

In Theorems 1 to 4, we have not assumed anything about the distribution of $f_i$. Now, let us consider an important special case where $f_i$ has infinite support. This assumption holds for many distributions used to model multi-path effect, e.g., Rayleigh, Rician and Nakagami fadings. Note that as $\gamma \to 0$ (equivalently, as delay goes to $\infty$), opportunism threshold $\kappa \to \infty$. Thus, from Theorem 4,

$$\lim_{\text{Delay} \to \infty} \left( \frac{E_b}{N_0} \right)_{\text{sys}}^{\Delta^*(\kappa)} = 0 \quad \text{w.p. 1.} \tag{20}$$

The relation (20) holds for any $\Phi(\cdot)$ with unbounded support. Though the minimal value is 0 in all the cases, the delay-energy tradeoff, i.e., how quickly energy approaches the minimum when the delay is increased, depends strongly on the distribution. It can be easily seen that if the tail of the distribution is heavy, then the delay increases at a smaller rate as $\kappa$ increases significantly with the increase in the delay. On the contrary, if the tail is lighter, the energy decreases at a slower rate with the increase in the delay. We explain this by considering a special class of distributions parametrized by $\alpha$. Fix $\alpha > 1$ and define $\Phi(x) = 0$ for $x < 1$ and $\Phi(x) = 1 - x^{(1-\alpha)}$ otherwise. Note that $\alpha$ determines how fast the tail of the distribution diminishes, e.g., as $\alpha$ increases the tail diminishes faster. For this distribution, clearly, delay $D = 1/\gamma = \kappa^{(\alpha-1)}$. Thus, when the power is within $O\left(\frac{1}{\kappa}\right)$ of the minimum, which is zero as $\Phi(\cdot)$ has unbounded support, the delay is $O(\kappa^{(\alpha-1)})$. If we pick $\alpha \in (1, 3/2)$, then the delay is $\omega(\sqrt{\kappa})$, i.e., the delay increases at rate strictly smaller than $\sqrt{\kappa}$. Contrast this with the delay-energy trade-off obtained in [21]. In [21], Neely considers a model in which the fast fading a Markov Chain (MC) with finite state space. In these settings, Neely shows that the delay is at least of the order of $\sqrt{V}$ when the energy is in the order $1/V$ neighborhood of minimum energy required for the stability. Note that the result holds for any transition probability matrix, and hence the steady state distribution of the MC that is used to model the fast fading. But, we have shown that the result of [21] does not hold when the fast fading distribution has $\infty$ support.

In the following result, we show that $\Delta^*(\kappa)$ becomes energy optimal as the delay goes to infinity. From the above discussion it should be clear that $\Delta^*(\kappa)$ becomes energy optimal as the delay goes to infinity when $\Phi(\cdot)$ has unbounded support. Hence, we only focus on the case where $\Phi(\cdot)$ is supported on a compact set. Let $B$ denote the supremum of the support.

*Theorem 5 (Energy Optimality):* Let the short term fading distribution $\Phi(\cdot)$ be supported on the interval $[0, B]$. As the number of users go to infinity, for every policy $\Delta$,

$$\left( \frac{E_b}{N_0} \right)_{\text{sys}}^{\Delta} \geq \frac{1}{B} \mathbb{E}_\Psi \left[ \frac{e^{\Gamma \Psi(S)}}{S} \right] \quad \text{w.p. 1.} \tag{21}$$

Moreover,

$$\lim_{\text{Delay}\to\infty} \left(\frac{E_b}{N_0}\right)_{\text{sys}}^{\Delta^*(\kappa)} \leq \frac{1}{B}\mathbb{E}_\Psi \left[\frac{e^{\Gamma\Psi(S)}}{S}\right] \quad \text{w.p. 1.} \qquad (22)$$

Thus, $\Delta^*(\kappa)$ becomes energy optimal as delay goes to $\infty$ in the mean field.

*Proof:* The relation (21) follows by observing that for user $i$, $d_i(t) \leq Bs_i$ for every $t$. Thus, the minimum sum power required to support rates $\Gamma/K$ for each user when fading is $Bs_i$ in every slot provides a lower bound on the sum power required to support the same rates with fading process $\vec{d}(t)$. Now, as shown in [1], in optimal power allocation in the multi-access channel is given by (1). Now, the relation (21) follows using the arguments similar to that in Theorem 3, where fading process is now given by $d_i(t) = Bs_i$ for every $i$.

Now, the relation (22) follows by taking the limit delay $\to \infty$ in (16). Note that as delay $\to \infty$, $\kappa \to B$, thus yielding (22). ∎

Caution has to be exercised while interpreting Theorem 5 as the limits are taken in two parameters, namely, the number of users $K \to \infty$ and the user delay $\to \infty$ (equivalently, $\kappa \to B$ or $\gamma \to 0$). Thus, the resulting limiting value depends on the relative rates at which these two parameters approach their respective limiting values. In Theorem 5, we first let $K \to \infty$ and subsequently, let $\gamma \to 0$. In other words, $K$ and $\gamma$ approach their respective limiting values, while $K\gamma$ is $\infty$. This can be clearly seen in (22) as we still see the superposition gain apparent in the term $e^{\Gamma\Psi(S)}$ as the optimal scheduling can take advantage of the variations in the pathloss values of different users. But, if we let $K \to \infty$ and $\gamma \to 0$, while ensuring $K\gamma \to 1$, then exactly one user will be scheduled under $\Delta^*(\kappa)$ for sufficiently large $K$. As a result, the superposition gain disappears. Thus, a question arises whether the energy optimality of $\Delta^*(\kappa)$ depends upon the relative rate at which $K$ and $\gamma$ approach their respective limits. Before answering this question, let us look at the lower bound (21). Note that this bound is tight if we split each of the $K$ users in $K$ different users (logical group) with the same pathloss, but i.i.d. short term fading, and let these $K$ users in a single logical group collectively desire the rate $\Gamma/K$, then let $K \to \infty$. Note that as $K \to \infty$, in each of the logical groups there exists a user with short term fading $B$ and scheduling such users from each of the logical groups at the rate $\Gamma/K$ simultaneously minimizes the sum power required to achieve the desired rates. Thus, in this setting, actual number of users, accounting for each user as $K$ different users, go to $\infty$ at rate $K^2$ and not at $K$. This shows that the tight lower bound also depends on the rate at which $K \to \infty$. Thus, for a fair comparison, one should scale the users in a logical group should scale at the rate $\gamma K$. With these scaling laws the optimality of $\Delta^*(\kappa)$ should hold. Indeed, it can be shown that in a special case $K\gamma = 1$, while $K \to \infty$ and $\gamma \to 0$, the lower and upper bounds both equal $\frac{(e^\Gamma-1)}{\Gamma B}\mathbb{E}[1/S]$ w.p. 1. Note that this value corresponds to scheduling a single user with the best short term fading value, which equals $B$ as $K \to \infty$, in every slot.

### A. Discussion on Analytical Results

Theorem 2 states that if the delay of $D \geq 1$ has to be provided, then the policy $\Delta^*(\kappa)$ can achieve it with any $\kappa$ satisfying $\mathbb{P}\{f_i > \kappa\} \geq 1/D$. Now, Theorem 4 states that choosing the largest $\kappa$ such that $\mathbb{P}\{f_i > \kappa\} = 1/D$ minimizes the system energy while providing the required delay when the number of users is large. Moreover, Theorem 3 quantifies the energy efficiency of the system for the given value $\kappa$. Note that Theorems 2, 3 and 4 together quantify the delay-energy tradeoff in the multi-access channel. Finally, Theorem 5 proves the energy optimality of $\Delta^*(\kappa)$ as the delay goes to $\infty$. Moreover, Theorem 4 shows that the decrease in energy efficiency with $\kappa$ is at least linear. Thus, given any value of energy $E$ greater than the minimum energy required to guarantee the rates to each of the users, there exists $\kappa$ such that $\Delta^*(\kappa)$ provides the desired rate to each user while maintaining the required sum energy below $E$.

### V. Generalizations

Now, we discuss two important generalizations. First, we consider the system with multiple non-overlapping bands. The required rate can be split on these bands. In Section V-A, we discuss how the results in Section IV can be generalized to this case. Second, we consider a case when the users need different delays. This is the case, when various types of applications are supported on multi-access channel, or when the multi-access channel serves as an intermediate hop on the multiple hops traveled by the application in the network. In Section V-B, we discuss how OSPC can support this.

### A. Multi-band Multi-access Channel

We consider multi-access channel with $M$ bands. We assume that the fading on these bands is statistically indistinguishable and independent. Let $f_i^m$ denote the short term fading for user $i$ on $m^{\text{th}}$ sub-band. Now, it is not immediately clear how the required rate should be split on the various bands in order to minimize the sum energy. But, fortunately, it has been shown that to minimize the sum energy required to realize a given rate vector on the multi-band multi-access channel, the total rate for a user should be supported on its best channel [30]. Let $f_i^*(t) \overset{\text{def}}{=} \max\{f_i^1(t) \ldots, f_i^M\}$. Thus, $\Delta^*(\kappa)$ has to be defined in terms $f_i^*(t)$ instead of $f_i(t)$, i.e., $\Delta^*(\kappa)$ selects all the users with $f_i^*(t) > \kappa$ and provides the required rate on the best channel for every user. Now, Theorems 1 and 2 hold with $\gamma \overset{\text{def}}{=} \mathbb{P}\{f_i^* > \kappa\}$. In Theorem 3, the energy efficiency becomes

$$\left(\frac{E_b}{N_0}\right)_{\text{sys}}^{\Delta^*(\kappa)} = \int_0^\infty \frac{1}{x} \exp\left(\frac{\Gamma}{M} \overline{F}_\kappa^*(x)\right) \mathrm{d}\overline{F}_\kappa^*(x) \quad \text{w.p. } 1,$$

where $\overline{F}_\kappa^*(\cdot)$ denote the fading distribution of user $i$ who is placed uniformly at random in a cell given that $f_i^*(t) > \kappa$. The additional factor of $1/M$ appears because only $1/M$ fraction of scheduled user transmit on a given band. Finally, Theorems 4 and 5 hold with $f_i$ replaced by $f_i^*$.

*B. Delay Differentiation*

The users are divided into $L$ classes based on their delay requirements. Let $\alpha_1, \ldots, \alpha_L$ fraction of users that want delays $D_1, \ldots, D_L$ respectively. Let $\kappa_l$ be the largest real number that satisfies $\mathbb{P}\{f_i > \kappa_l\} = \frac{1}{D_l}$ for every $l \leq L$. Now, OSPC can use $\kappa_l$ instead of $\kappa$ for users of class $l$. Clearly, Theorems 1 and 2 hold with $\gamma_l \stackrel{\text{def}}{=} \mathbb{P}\{f_i > \kappa_l\}$ for every $l$. Moreover, energy efficiency of each class $l$ can be computed along the similar lines as the proof of Theorem 3. Now, the energy efficiency for the system is the weighted sum (with respect to $\alpha_l$'s) of the energy efficiency of each class. Finally, Theorems 4 and 5 can be shown to hold for each class individually. Since, the system energy efficiency is the convex combination of the energy efficiencies of the classes, Theorems 4 and 5 follow for the whole system.

## VI. Numerical Results

We consider a system where users are placed uniformly at random in a cell except for a forbidden region around the access point of radius $\delta = 0.01$. The path loss exponent is two ($\alpha = 2$). All users experience short term fading with exponential energy distribution with mean one on each of the ten ($M = 10$) independently fading bands. The explicit mathematical formulations for the channel models can be found in Appendix D. The path loss model is normalized to unity at cell edge, so that the results should be normalized with a corresponding factor. This, however, has no effect on the relative numerical results we report.

Fig. 1 demonstrates the delay energy tradeoff exhibited by OSPC. For an increase in average delay from one to three slots, an energy saving of over 3dB is gained. Thus, even the small delay tolerance of the application can be exploited to obtain significant improvement in the energy efficiency of the system. Moreover, the energy required to support the given rate decreases monotonically as delay increases. This verifies Theorem 4. Note that the gain exhibits very similar behavior across various spectral efficiencies.

Fig. 2 provides a comparison between OSPC and PFS. The mathematical expressions for computing the energy efficiency ($E_b/N_0$) under PFS can be found in Theorem 1 of [30] and are given in Appendix E for completeness. In the delay-limited case with the strict delay constraint of a single slot (i.e. $\kappa = 0$), OSPC can outperform PFS only at high spectral efficiencies. However, as delay tolerance of the application increases, OSPC can outperform PFS over a wider range of spectral efficiencies. Also, the improvement in the energy efficiency under OSPC over that under PFS increases monotonically with increase in the delay tolerance. Note that the improvement happens while guaranteeing a required rate for each user, which is not the case for PFS. A notable feature of OSPC is that changing the opportunism threshold results in a horizontal shift of the performance curve, which again indicates the energy-delay tradeoff behaves in a similar manner for all system spectral efficiencies.

An empirical verification for the convergence of the system energy can be found in Figure 3, where the smallest and largest empirically found energy efficiencies of an ensemble of 1000 simulated systems are reported. Each system has a different random pathloss vector. The system employs hard fairness, i.e. it does not allow for delay, and data is assumed to arrive at users' transmit buffers in each slot. The extreme energies converge towards the asymptotic at $K = \infty$ as the user population is increased.

Figure 3 indicates that the system energy efficiency deviates from asymptotic behavior at high spectral efficiency and with a smaller number of users. This behavior is due to the following. With a finite number of users, the transmitted rate is no longer deterministic as in the asymptotic case. Instead, the number of simultaneously scheduled users and their buffer lengths vary from slot to slot. This results in energy loss through the convexity of the exponential rate-energy function in 1. Since the rate-energy function has the spectral efficiency as a multiplicative factor in the exponential, the loss is greater at high spectral efficiency.

## VII. Conclusions

We showed that by opportunistically choosing a suitable fraction of users with the best channels in each slot, we can improve the energy efficiency of the system while providing the required delay to each user. Since the policy empties the scheduled users' queues, it has good stability properties. We showed that the expected user delay is inversely proportional to the scheduling fraction. Delay can then be adjusted simply by choosing an appropriate opportunism threshold, while delay differentiation can be achieved by applying different thresholds for different delay classes. Moreover, if the application does not need any delay guarantees, then OSPC can achieve any required energy efficiency ($E_b/N_0 > 0$) while maintaining system stability. The scheme performs well compared to PFS, while providing rate guarantees.

## VIII. Acknowledgements

The work of P. Chaporkar was supported by ERCIM Fellowship and the work of K. Kansanen and R. R. Müller by the Research Council of Norway (NFR) under the NORDITE/VERDIKT programme (NFR contract no. 172177). The authors wish to acknowledge Mr. Majid Butt for providing some of the numerical results. Parts of this paper were presented in The 3rd workshop on Resource Allocation in Wireless Networks (RAWNER 2007), April 16th, Limassol, Cyprus.

# APPENDIX A

## PROOF OF LEMMA 1

*Proof:* First, we show that for every $x$ and $t$

$$\lim_{K \to \infty} \sum_{k \in \mathcal{A}_K^\kappa(x,t)} \rho_k^{\Delta^*(\kappa)}(t) = \Gamma \overline{F}_\kappa(x) \quad \text{w.p. 1.} \tag{23}$$

If $x < \kappa$, then the result trivially holds as both sides of (23) are equal to zero. Hence, we consider $x > \kappa$ in the following.

For a chosen user $k$, the rate $\rho_k^{\Delta^*(\kappa)}$ is equal to its total buffer occupancy in slot $t$. Under $\Delta^*(\kappa)$, buffer occupancy in slot $t$ is equal to the number arrivals since the last time $k$ was scheduled. Let $\overline{\nu}_k \stackrel{\text{def}}{=} \sum_{u=\tau(t)+1}^{t} \nu_k(u)$, where $\tau(t) = \max\{u < t : f_k(u) > \kappa\}$. Note that $\nu_k(u)$ and the fading are i.i.d. across both the slots and users, and they are also mutually independent. So, clearly, $\overline{\nu}_k$ are i.i.d. across the chosen users and $\mathbb{E}[(\overline{\nu}_k)] = \frac{1}{\gamma} \mathbb{E}[(\nu_k)] = 1/\gamma$. Moreover, for a chosen user $k$

$$\rho_k^{\Delta^*(\kappa)}(t) = \frac{\Gamma}{K} \overline{\nu}_k. \tag{24}$$

Note that $|\mathcal{A}_K^\kappa(x,t)| \to \infty$ as $K \to \infty$ for every $t$ and $x > \kappa$.

$$\begin{aligned}
&\lim_{K \to \infty} \sum_{k \in \mathcal{A}_K^\kappa(x,t)} \rho_k^{\Delta^*(\kappa)}(t) \\
&= \lim_{K \to \infty} \sum_{k \in \mathcal{A}_K^\kappa(x,t)} \frac{\Gamma}{K} \overline{\nu}_k \quad \text{(from (24))} \\
&= \Gamma \lim_{K \to \infty} \frac{|\mathcal{A}_K^\kappa(\infty,t)|}{K} \lim_{|\mathcal{A}_K^\kappa(\infty,t)| \to \infty} \frac{|\mathcal{A}_K^\kappa(x,t)|}{|\mathcal{A}_K^\kappa(\infty,t)|} \lim_{|\mathcal{A}_K^\kappa(x,t)| \to \infty} \frac{\sum_{k \in \mathcal{A}_K^\kappa(x,t)} \overline{\nu}_k}{|\mathcal{A}_K^\kappa(x,t)|} \\
&= \Gamma \overline{F}_\kappa(x) \quad \text{w.p. 1.}
\end{aligned} \tag{25}$$

The relation (25) follows as fading and arrivals are independent. Note that the first limit converge to $\gamma$, the second to $\overline{F}_\kappa(x)$ and the last one to $1/\gamma$.

Now, we claim that the channel gains $d_k$'s for the chosen users can be viewed as i.i.d. variables. Note that if $\Delta^*(\kappa)$ had scheduled users based on $d_k$ rather than on $f_k$, then $d_k$ for the chosen users would not be i.i.d. as the users that are nearer to the receiver are likely to be favored. Since $f_k$'s are i.i.d. irrespective of the distance between the user and receiver, the scheduling decision can be viewed as scheduling each user w.p. $\gamma$ independently. Since the users are placed at random, $s_k$ is a deterministic function of distance, and $f_k$ and $s_k$ are independent, we conclude that $d_k$'s for the chosen users are i.i.d. and each $d_k$ is distributed as $\overline{F}_\kappa(\cdot)$. Finally, (7) follows from (25) using Glivenko-Catelli Theorem. ∎

## APPENDIX B
### PROOF OF LEMMA 2

*Proof:* For simplicity, we prove the required when $\nu_i$ has finite support, say $\nu_{\max}$. Let $\tau_0(i)$ denote last time before $t$ user $i$ was scheduled, i.e., $\tau_0(i) = \max_{u<t}\{f_i(u) > \kappa\}$. Moreover, let $T_i = t - \tau_0(i)$ for each user $i$. Note that $\{T_i\}_{i\in\{1,...,K\}}$ is i.i.d. sequence and $\mathbb{P}\{T_i > u\} = (1-\gamma)^u$ for every $i$.

Fix any user $i \in \mathcal{A}_K^\kappa(x,t)$, and observe that

$$\rho_i^{\Delta^*(\kappa)}(t) = \frac{\Gamma}{K}\sum_{\tau_0(i)+1}^{t}\nu_i(t) \leq \frac{\Gamma\nu_{\max}T_i}{K}. \qquad (26)$$

Next, fix $\epsilon > 0$, and consider

$$
\begin{aligned}
\mathbb{P}\left\{\sup_{i\in\mathcal{A}_K^\kappa(x,t)}\rho_i^{\Delta^*(\kappa)}(t) > \epsilon\right\} &\leq \sum_{i\in\mathcal{A}_K^\kappa(x,t)}\mathbb{P}\{\rho_i^{\Delta^*(\kappa)}(t) > \epsilon\} \quad \text{(by Union bound)} \\
&\leq \sum_{i\in\mathcal{A}_K^\kappa(x,t)}\mathbb{P}\left\{\frac{\Gamma\nu_{\max}T_i}{K} > \epsilon\right\} \quad \text{(by (26))} \\
&\leq \sum_{i=1}^{K}\mathbb{P}\left\{\frac{\Gamma\nu_{\max}T_i}{K} > \epsilon\right\} \quad \text{(as $|\mathcal{A}_K^\kappa(x,t)| \leq K$ for all $x$ and $t$)} \\
&= K\mathbb{P}\left\{T_i > \frac{\epsilon K}{\nu_{\max}\Gamma}\right\} \quad \text{(as $T_i$'s are i.i.d.)} \\
&= K(1-\gamma)^{\frac{\epsilon K}{\Gamma\nu}}.
\end{aligned}
$$

Since $\gamma > 0$, it follows that

$$\sum_{K=1}^{\infty}\mathbb{P}\left\{\sup_{i\in\mathcal{A}_K^\kappa(x,t)}\rho_i^{\Delta^*(\kappa)}(t) > \epsilon\right\} \leq \sum_{K=1}^{\infty}K(1-\gamma)^{\frac{\epsilon K}{\Gamma\nu}} < \infty.$$

Now, the result follows by Borel-Cantelli Theorem as $\epsilon$ was arbitrary. ∎

APPENDIX C

PROOF OF LEMMA 3

*Proof:* The proof is by construction. We construct a sequence of rate vectors $\{\vec{\rho}(u)\}_{u \geq 0}$ such that $\vec{\rho}(0) = \vec{\rho}$ and $\lim_{u \to \infty} \vec{\rho}(u) = \vec{\rho}'$. Let $\vec{E}(u)$ denote the power allocation to realize $\vec{\rho}(u)$. Then, we show that $\sum_{i \in \mathcal{A}} E_i(u)$ is a non-increasing function of $u$, and thus proving the required. The recursive procedure to construct $\vec{\rho}(u)$ is as follows.

Initialize: $\vec{\rho}(0) = \vec{\rho}$.

STEP $u$:

C1. $\rho_i(u) = \rho_i(u-1)$ for every $i \notin \{u, u+1\}$,

C2. $\rho_u(u) = \rho'_u$,

C3. $\rho_{u+1}(u) = \rho_{u+1}(u-1) + \rho_u(u-1) - \rho'_u$.

Note that $\vec{\rho}(u)$ satisfies $\rho_i(u) = \rho'_i$ for every $i \leq u$, and $\rho_i(u) = \rho_i$ for every $i > u+1$. Thus, clearly, $\lim_{u \to \infty} \vec{\rho}(u) = \vec{\rho}'$. Now, using induction, we prove that for every $u$

$$\sum_{i=1}^{u+1} \rho_i(u) = \sum_{i=1}^{u+1} \rho_i. \tag{27}$$

Note that (27) holds for $u = 0$ as $\rho_1(0) = \rho_1$ by initialization. Now, let (27) hold for every $u = 0, \ldots, n-1$. We show (27) for $u = n$.

Since (27) holds for $u = n-1$, we know that

$$
\begin{aligned}
\sum_{i=1}^{n} \rho_i &= \sum_{i=1}^{n} \rho_i(n-1) \\
&= \sum_{i=1}^{n-1} \rho_i(n) + \rho_n(n-1) \quad \text{(by STEP C1.)} \\
&= \sum_{i=1}^{n+1} \rho_i(n) - \rho_{n+1}(n-1) \quad \text{(from step C3.)} \\
&= \sum_{i=1}^{n+1} \rho_i(n) - \rho_{n+1} \quad \text{(as } \rho_i(u) = \rho_i \text{ for every } i > u+1).
\end{aligned}
$$

This proves (27) for every $u$.

Now, we show that $\sum_{i \in \mathcal{A}} E_i(u)$ is non-increasing function of $u$. From (1) and (27), it is clear that $E_i(u) = E_i(u-1)$ for every $i \notin \{u, u+1\}$. Thus it suffices to consider

$$
\begin{aligned}
&[E_u(u) + E_{u+1}(u)] - [E_{u-1}(u) - E_{u+1}(u-1)] \\
&= \frac{N_0}{d_u} \left[ e^{\sum_i^u \rho_i(u)} - e^{\sum_i^u \rho_i(u-1)} \right] - \frac{N_0}{d_{u+1}} \left[ e^{\sum_i^u \rho_i(u)} - e^{\sum_i^u \rho_i(u-1)} \right] \\
&= N_0 \left[ \frac{1}{d_u} - \frac{1}{d_{u+1}} \right] \left[ e^{\sum_i^u \rho_i(u)} - e^{\sum_i^u \rho_i(u-1)} \right]
\end{aligned}
$$

$$\begin{aligned} &= N_0 \left[ \frac{1}{d_u} - \frac{1}{d_{u+1}} \right] \left[ e^{\sum_i^u \rho_i'} - e^{\sum_i^u \rho_i} \right] \quad \text{(by STEPs C1 and C2, and (27))} \\ &\leq 0. \end{aligned}$$

The last inequality follows as $d_u \leq d_{u+1}$ and $\sum_i^u \rho_i' \leq \sum_i^u \rho_i$ by suppositions in the lemma.

Now, to prove the required, we need to show that $\lim_{u \to \infty} \sum_{i \in \mathcal{A}} E_i(u) = \sum_{i \in \mathcal{A}} E_i'$. First, note that if $\sum_{i \in \mathcal{A}} E_i = \infty$, then the result immediately follows. So, we consider a non-trivial case, $\sum_{i \in \mathcal{A}} E_i < \infty$. We know that for every $u$, $0 \leq \sum_{i \in \mathcal{A}} E_i(u) \leq \sum_{i \in \mathcal{A}} E_i$. Moreover, the sequence $\left\{ \sum_{i \in \mathcal{A}} E_i(u) \right\}_{u \geq 0}$ is monotone. Thus, $\lim_{u \to \infty} \sum_{i \in \mathcal{A}} E_i(u)$ exists. Moreover, by dominated convergence theorem, we can exchange the limit and summation. Thus,

$$\begin{aligned} \lim_{u \to \infty} \sum_{i \in \mathcal{A}} E_i(u) &= \sum_{i \in \mathcal{A}} \lim_{u \to \infty} E_i(u) \\ &= \sum_{i \in \mathcal{A}} \lim_{u \to \infty} \left[ \frac{N_0}{d_i} e^{\sum_{k=1}^{i-1} \rho_k(u)} (e^{\rho_i(u)} - 1) \right] \quad \text{(from (1))} \\ &= \sum_{i \in \mathcal{A}} \left[ \frac{N_0}{d_i} e^{\sum_{k=1}^{i-1} \rho_k'} (e^{\rho_i'} - 1) \right] \quad \text{(as } \vec{\rho}(u) \to \vec{\rho}' \text{ and } \exp(\cdot) \text{ is continuous)} \\ &= \sum_{i \in \mathcal{A}} E_i'. \end{aligned}$$

This proves the required. ∎

# APPENDIX D

## CHANNEL STATISTICS

Channel state is assumed to remain constant during one transmission slot and assume a new value for each slot independently for all slots and users. We, thus, avoid the dependence on users and time in the following.

Users are placed uniformly on a circular cell. The channel state of each user is the product of two independent ergodic random processes, path loss and short term fading. Path loss is exponentially dependent on the distance of the user from the access point, and is assumed to remain constant for a users across transmission slots. The distance dependency is parametrized by the path loss exponent $\alpha$, usually ranging within the interval $[2, 4]$. To avoid a singularity for users next to the access point, a forbidden circular region of radius $\delta$ is created around the access point. This model results in the following cumulative distribution of the path loss

$$P_s(x) = \begin{cases} 0 & x \leq 1 \\ 1 - \frac{x^{-2/\alpha} - \delta^2}{1 - \delta^2} & 1 \leq x \leq \delta^{-\alpha} \\ 1 & x \geq \delta^{-\alpha} \end{cases} . \tag{28}$$

Note that the model is normalized to provide unit path loss at cell border. Thus, the results reported here must be rescaled in terms of $(E_b/N_0)_{\text{sys}}$ by a factor of $D_0^{-\alpha}$, where $D_0$ denotes the actual radius of the cell. Naturally, since all results behave accordingly, all comparisons remain valid.

The short term fading process on each band $m$ is modeled by a zero-mean Gaussian circularly symmetric random variable (i.e. Rayleigh fading), with an exponentially distributed envelope. It assumes a new value for each user in each slot. The multiple bands are assumed independently fading. In each slot, those users $i$ whose maximum channel gain (over all $M$ channels) $\max\left\{f_i^m(t)\dots f_i^M(t)\right\} > \kappa$ are scheduled by OSPC on each band, and the cumulative short term fading distribution is given by

$$P_f(x) = 1 - \frac{1 - (1 - e^{-x})^M}{\gamma}, x \in [\kappa, \infty), \tag{29}$$

where $\gamma = \mathbb{P}\left(f_i > x\right) = 1 - (1 - e^{-\kappa})^M$. With some algebra, the cumulative distribution for the (product) channel can be expressed as

$$\overline{F}_\kappa^*(x) = \begin{cases} \frac{1 - \kappa/x}{1 - \delta^2} - \frac{1}{\gamma x (1 - \delta^2)} \sum_{i=1}^M \frac{1}{i} \left[ (1 - \exp(-x))^i - (1 - \exp(-\kappa))^i \right] & x < \kappa \delta^{-2} \\ 1 - \frac{1}{\gamma x (1 - \delta^2)} \sum_{i=1}^M \frac{1}{i} \left[ (1 - \exp(-x))^i - \left(1 - \exp(-x\delta^2)\right)^i \right] & x \geq \kappa \delta^{-2} \end{cases}. \tag{30}$$

## Appendix E

## Proportional Fair Scheduling

We follow completely the approach of [30] in the evaluation. The average spectral efficiency of the system with PFS is given implicitly by [30]

$$\mathsf{C} = \int_0^\infty \log_2\left(1 + x\mathsf{SNR}\right) dF_{s\max\{f\}, K}(x) \tag{31}$$

$$\left(\frac{E_b}{N_0}\right)_{\mathrm{sys}}^{\mathrm{PFS}} = = \frac{\mathsf{SNR}}{\mathsf{C}} \tag{32}$$

where $F_{s\max\{f\}, K}(x)$ denotes the distribution of the product of the random path loss and the maximum of $K$ users' short term fading coefficients, $s\max\{f_k, \dots, f_K\}$, and is given by

$$F_{s\max\{f\}, K}(x) = 1 - \frac{1}{x\left(1 - \delta^2\right)} \sum_{i=1}^K \frac{1}{i} \left[ (1 - \exp(-x))^i - \left(1 - \exp(-x\delta^2)\right)^i \right] \tag{33}$$

which is identical to (30) when $\gamma = 1$, $\kappa = 0$ and $M = K$.

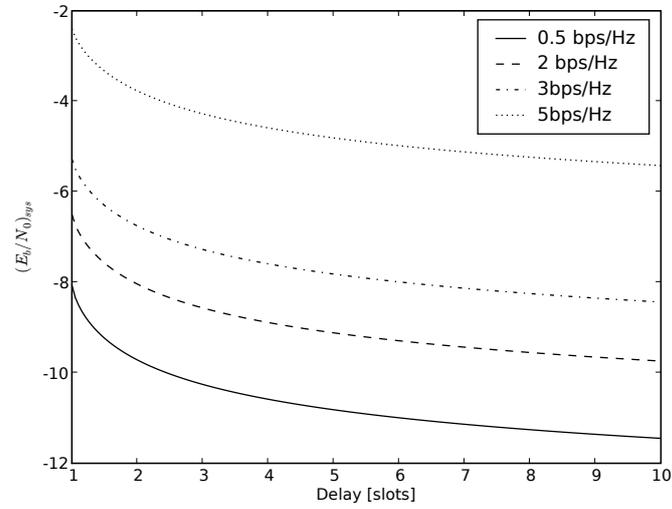

Fig. 1.  System $E_b/N_0$ as a function of delay

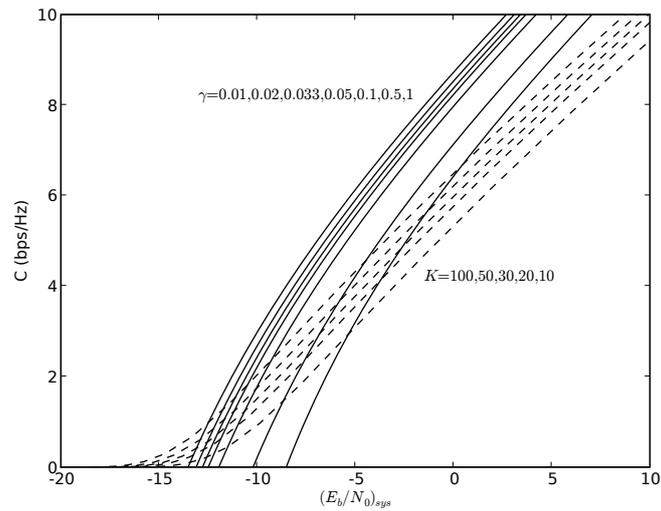

Fig. 2.  Comparison between PFS (dashed) and OSPC (solid line).

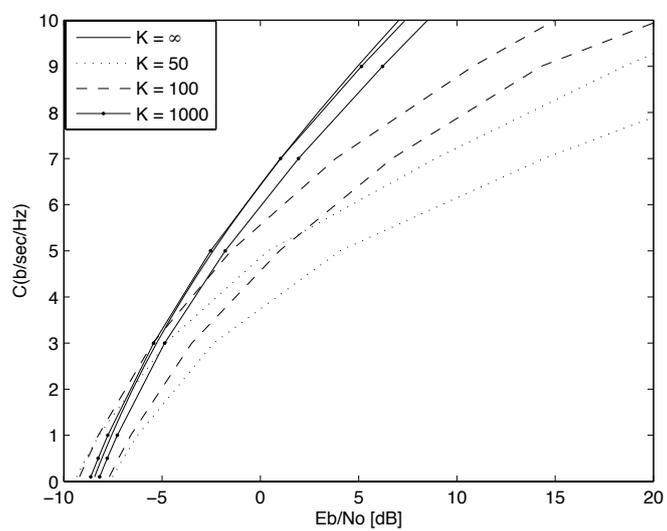

Fig. 3.   Empirical convergence of extremes with a finite number of users